\documentclass{emulateapj}
\usepackage{epsf}
\usepackage{times}
\newcommand{\etal}{{et al}\/.}

\begin{document}
\slugcomment{ApJ Letters, accepted 5th Oct 2007}
\shorttitle{Particle acceleration in Cen A}
\shortauthors{M.J.\ Hardcastle \etal}
\title{New results on particle acceleration in the Centaurus A jet and
  counterjet from a deep {\it Chandra} observation}
\author{M.J.\ Hardcastle\altaffilmark{1}, R.P.\ Kraft\altaffilmark{2}, G.R.\
  Sivakoff\altaffilmark{3}, J.L.\
  Goodger\altaffilmark{1}, J.H.\ Croston\altaffilmark{1}, 
  A.\ Jord\'an\altaffilmark{2,4}, D.A.\ Evans\altaffilmark{2}, D.M.\ Worrall\altaffilmark{5,2}, M.\ Birkinshaw\altaffilmark{5,2},  S.\
  Raychaudhury\altaffilmark{6,2}, N.J.\ Brassington\altaffilmark{2}, W.R.\ Forman\altaffilmark{2}, W.E.\ Harris\altaffilmark{7}, C.\ Jones\altaffilmark{2}, A.M.\ Juett\altaffilmark{8}, S.S.\ Murray\altaffilmark{2}, P.E.J.\ Nulsen\altaffilmark{2}, C.L.\
  Sarazin\altaffilmark{8}, and K.A.\ Woodley\altaffilmark{7}}
\altaffiltext{1}{School of Physics, Astronomy \& Mathematics, University of
  Hertfordshire, College Lane, Hatfield AL10 9AB, UK;
  m.j.hardcastle@herts.ac.uk, j.l.goodger@herts.ac.uk, j.h.croston@herts.ac.uk}
\altaffiltext{2}{Harvard-Smithsonian Center for Astrophysics, 60
  Garden Street, Cambridge, MA~02138, USA; kraft@head.cfa.harvard.edu,
  ajordan@cfa.harvard.edu, devans@cfa.harvard.edu, nbrassington@cfa.harvard.edu,
  wforman@cfa.harvard.edu, cjones@cfa.harvard.edu,
  ssm@cfa.harvard.edu, pnulsen@cfa.harvard.edu}
\altaffiltext{3}{Department of Astronomy, The Ohio State University,
  4055 McPherson Laboratory 140 W. 18th Avenue, Columbus, OH
  43210-1173, USA; sivakoff@astronomy.ohio-state.edu}
\altaffiltext{4}{European Southern Observatory,
  Karl-Schwarzschild-Str.\ 2 85748 Garching bei M\"{u}nchen, Germany}
\altaffiltext{5}{Department of Physics, University of Bristol, Tyndall
  Avenue, Bristol BS8 ITL, UK; d.m.worrall@bristol.ac.uk, mark.birkinshaw@bristol.ac.uk}
\altaffiltext{6}{School of Physics and Astronomy, University of
  Birmingham, Edgbaston, Birmingham B15 2TT, UK; somak@star.sr.bham.ac.uk}
\altaffiltext{7}{Department of Physics and Astronomy, McMaster
  University, Hamilton, ON L8S 4M1, Canada;
  harris@physics.mcmaster.ca, woodleka@physics.mcmaster.ca}
\altaffiltext{8}{Department of Astronomy, University of Virginia, P.
  O. Box 400325, Charlottesville, VA 22904-4325, USA;
  ajuett@virginia.edu, sarazin@virginia.edu}
\begin{abstract}
We present new deep {\it Chandra} observations of the Centaurus A jet,
with a combined on-source exposure time of 719 ks. These data allow
detailed X-ray spectral measurements to be made along the jet out to
its disappearance at 4.5 kpc from the nucleus. We distinguish several
regimes of high-energy particle acceleration: while the inner part of
the jet is dominated by knots and has properties consistent with local
particle acceleration at shocks, the particle acceleration in the
outer 3.4 kpc of the jet is likely to be dominated by an unknown
distributed acceleration mechanism. In addition to several compact
counterjet features we detect probable extended emission from a
counterjet out to 2.0 kpc from the nucleus, and argue that this
implies that the diffuse acceleration process operates in the
counterjet as well. A preliminary search for X-ray variability finds
no jet knots with dramatic flux density variations, unlike the
situation seen in M87.
\end{abstract}

\section{Introduction}
\label{intro}
Low-power, Fanaroff-Riley class I (FRI) \citep{fr74} radio galaxies
are numerically the dominant population of
radio-loud active galaxies in the universe. Their dynamics and energy
content are thus essential to models of AGN `feedback' in which the
energy released in the process of accretion onto the central
supermassive black hole is transported to large spatial scales via the
interaction between the jets and the external medium. To understand
the dynamics and energetics of these sources it is crucial that we
should understand the particle acceleration processes by which the
bulk kinetic energy of the jets is translated into the internal energy
of relativistic plasma.

The discovery with {\it Chandra} that X-ray emission is common in the
inner few kpc of FRI jets \citep{wbh01} provides a very strong
argument that {\it in situ} high-energy particle acceleration is
taking place in those regions. The broad-band spectral energy
distribution and X-ray spectrum of the X-ray emission imply a
synchrotron origin for the X-rays \citep{hbw01}. For magnetic fields
close to the equipartition value in a typical powerful jet, the loss
timescale ($\tau = -E/\dot E$) for the $\gamma \sim 10^7$ -- $10^8$
electrons emitting X-ray synchrotron emission is tens of years. Thus
observations of X-ray synchrotron emission essentially tell us where
particle acceleration is happening {\it now}: in particular, resolved,
diffuse X-ray emission implies a particle acceleration
mechanism that must be distributed throughout the jet.

To probe the nature of the acceleration mechanisms in FRI jets we
require observations that can reliably distinguish between compact and
diffuse X-ray emission: this implies (at the most optimistic) a
spatial resolution that is comparable to the loss spatial scale, the
distance travelled by an electron before synchrotron losses remove it
from the X-ray band, or roughly $c\tau$. Even with the comparatively
high angular resolution of {\it Chandra}, this can only be achieved in
the nearest FRI radio galaxy, Centaurus A. Given Cen A's distance of
$\sim 3.7$ Mpc (the average of 5 distance estimates to Cen A:
\citealt{fmst07}), {\it Chandra}'s resolution corresponds to $\sim 10$
pc.

{\it Chandra} observations of Cen A show a complex, knotty and in
places edge-brightened jet structure (\citealt{kfjk00, kfjm02, hwkf03}
[hereafter H03]; \citealt{ksat06} [K06]; \citealt{hkw06}
[H06]). It has been argued (H03,K06) that at least two
acceleration mechanisms are required: the dynamics and spectra of the
compact knots in the inner part of the jet are consistent with a shock
origin, while the diffuse emission is inconsistent both spectrally
(H03) and in terms of the number distribution of knots (K06) with
being the sum of many unresolved knots with the same properties as
those observed, and is instead probably truly diffuse, implying a
distributed acceleration mechanism such as second-order Fermi
acceleration \citep[e.g.,][]{so02} or magnetic field line reconnection
\citep[e.g.,][]{bl00}. Here we present the results of new, much deeper
{\it Chandra} observations and their consequences for the nature of
particle acceleration in the Cen A jet and counterjet.

\begin{figure*}
%\plotone{/home/mjh/pictures/CENA-JETCOL.PS}
\epsfxsize 16.6cm
\epsfbox{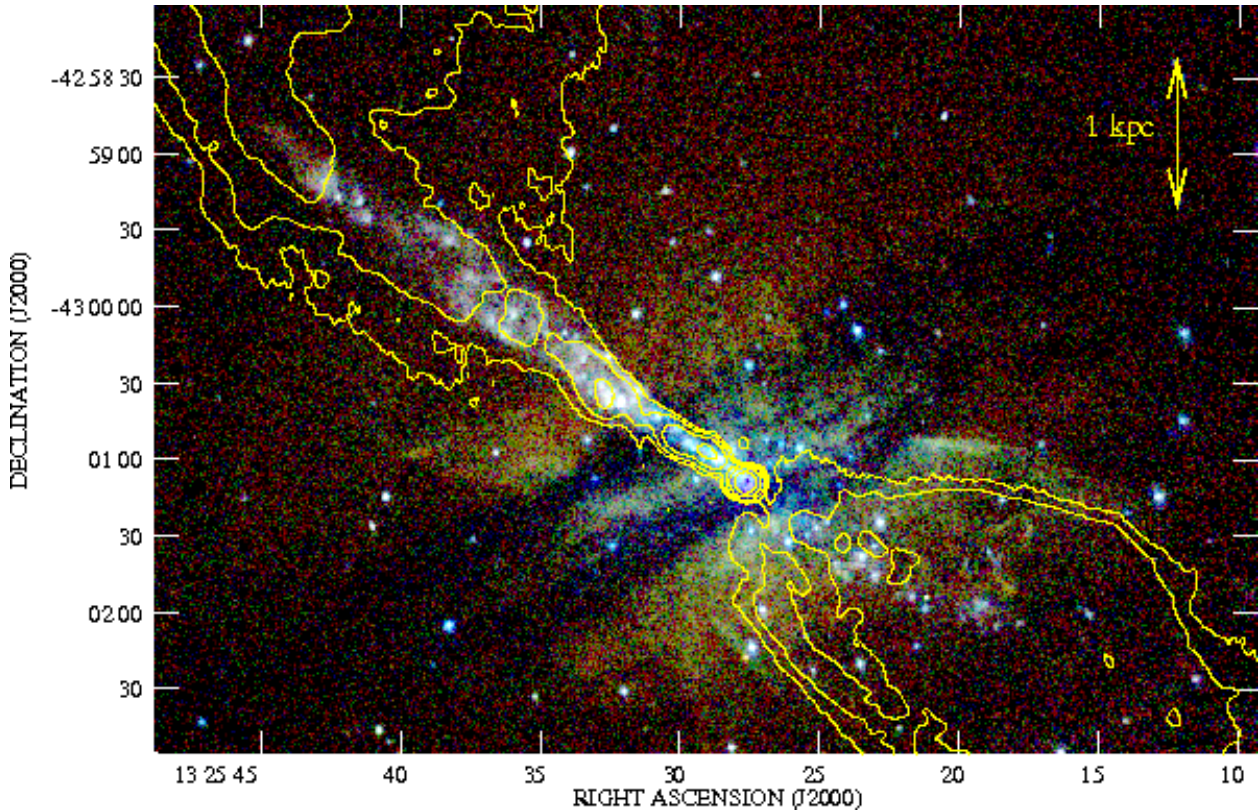}
%\vskip -10pt
\caption{False-color image of the jet and counterjet regions of Cen A
  using all ACIS observations. The {\it Chandra} data from each
  observation have been exposure corrected at an appropriate energy
  and then combined, weighting according to the value of the exposure
  map. The images are binned in standard {\it Chandra} pixels
  ($0\farcs492$ on a side) and smoothed with a FWHM = $1\farcs0$
  Gaussian. Red shows exposure-corrected counts in the energy range
  0.4--0.85 keV, green shows 0.85--1.3 keV and blue 1.3--2.5 keV.
  Contours [$7 \times (1,4,16\dots)$ mJy beam$^{-1}$] are from the
  5-GHz VLA map of H06 with $6\farcs0$ resolution.}
\label{fc}
\end{figure*}

In this analysis {\it Chandra} data processing was done using {\sc
ciao} 3.4 and {\sc caldb} 3.3.0.1. Spectral fitting was carried out in
{\sc xspec}. We define the spectral index $\alpha$ such that flux
$\propto \nu^{-\alpha}$: the photon index $\Gamma = 1+\alpha$. Errors
quoted and plotted throughout are $1\sigma$ for one interesting parameter.

\section{Observations}

The {\it Chandra} AXAF CCD Imaging Spectrometer (ACIS) has observed
Cen A four times prior to 2007 (OBSID 316, 962, 2978 and 3965)
for a total of 157 ks (see e.g., \citealt{kfjm02}; H03; K06) and six
additional times in 2007 for the Cen A Very Large Project (Cen
A-VLP; OBSID 7797, 7798, 7799, 7800, 8489 and 8490). The
combined ACIS data set has an effective on-source time of 719 ks. All
X-ray spectral fits are the result of joint fitting to spectra
extracted from matching regions from each separate observation.  Fig.\
\ref{fc} shows a false-color image of the jet and counterjet region.

Cen A has also been extensively observed in the radio using the NRAO
Very Large Array (VLA). In this paper we make use of radio data whose
reduction was described by H03 and H06. We also use new
observations at 8.4 GHz taken on 2007 Jun 04 (observation AG754) as
part of our ongoing program to monitor proper motions and variability
in the Cen A radio jet. These data and the results of the monitoring
program will be presented elsewhere (Goodger \etal , in preparation).
For radio/X-ray comparisons, we have aligned both the radio and X-ray
frames to the VLBI position of the nucleus quoted by \cite{maef98}.
The relative astrometric alignment of the {\it Chandra} data was done
in the manner described by \cite{jsmb07}. We then
determined the X-ray centroid of the nucleus by two independent
methods (determining the intersection of the readout streaks and
centroiding on the hard wings of the AGN PSF) which gave consistent
results. The resulting radio/X-ray alignment is accurate to $\pm
0\farcs1$.

Fig.\ \ref{fc} shows a number of interesting features of the new deep
{\it Chandra} observation. The jet is now detected with high signal to
noise out to $250''$ (4.5 kpc in projection) from the active nucleus,
and this allows us to extract spectra for a large number of jet
regions; we discuss the jet spectra in \S\ref{spectra}. Diffuse
emission plausibly related to a counterjet is now detected out to
$110''$ (2.0 kpc); we discuss the nature of this emission and its
implications in \S\ref{cjet-section}. Constraints on large-scale
variability in jet knots are discussed in \S\ref{variability}.

\section{Spectra of diffuse and compact features}
\label{spectra}

To investigate the jet X-ray spectrum we followed K06 in dividing the
jet into compact and diffuse regions. We defined as a knot any compact
feature in the jet that is clearly distinguished (by a factor $\ga 2$)
in surface brightness from its surroundings and has radius $<2''$.
These criteria include many, but not all, of the knots discussed by
\citet{kfjm02} and K06: by our definition there are 26 knots in the
jet, all but 4 of which are bright enough for spectral analysis. We
extracted spectra for these knots using the {\sc psextract} script,
fitting a power-law spectrum with a single normalization and photon
index and with absorption that was free to vary above the Galactic
value (we adopt $N_{\rm H} = 8.4 \times 10^{20}$ cm$^{-2}$, as
reported by the {\sc colden} code: \citealt{dl90}), but constrained
not to fall below it. The effects of knot variability on
this fitting are discussed in \S\ref{variability}. Background
subtraction was done using local off-source background regions, taking
account of the variation in both background surface brightness and
column density. The knot spectra may therefore include a small amount
of contamination from diffuse jet background.

The diffuse region of the jet was taken to be a polygonal region
including all jet-related emission but excluding all features defined
above as knots. We divided this region into rectangular sub-regions
whose long axes were the whole transverse width of the jet (transverse
spectral variations in the jet are discussed by \citealt{wbks07}) and
whose short axes were allowed to grow in a direction parallel to the
jet axis, in steps of one {\it Chandra} pixel, until sufficient counts
were enclosed in all regions for spectroscopic analysis ($\ga 1000$
counts in the combined dataset before background subtraction). This
process gave a total of 53 extended regions. Spectra for these regions
were extracted with the {\sc specextract} script and were fitted with
absorbed power-law models in precisely the same manner as the knots.
All the fits for knots and extended regions were acceptable
(probability under the null hypothesis $>2$\%).

\begin{figure}
\epsfxsize 8.6cm
\epsfbox{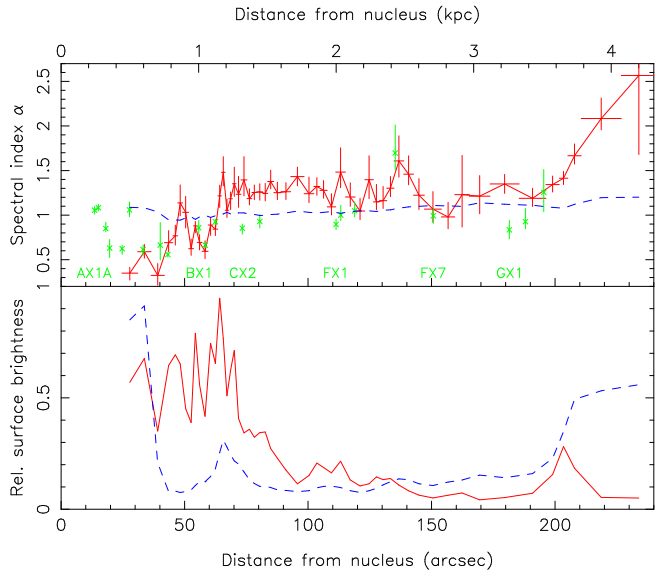}
\caption{Spectral index and surface brightness in the jet of Cen A.
  Top panel: red line shows the spectral index in the extended
  regions ($\alpha_{\rm X}$), dashed blue line shows the
  radio/X-ray two-point spectral index, $\alpha_{\rm RX}$, for
  extended emission. Green stars show the positions and X-ray
  spectral indices of knots. Vertical bars are $1\sigma$ errors:
  horizonal bars show the size of extended extraction regions.
  Bottom panel: surface brightness profile (arbitrary units) for
  extended emission in X-ray (red solid line) and radio (blue
  dashed line).}
\label{profile}
\end{figure}
The results of this process are summarized in Fig.\ \ref{profile},
which may be compared with fig.\ 4 of K06 and table 2 of
H03. Based on this profile, the extended
emission of the jet can be divided into three regions with different
radio and X-ray properties:

\noindent --- (i) ($\theta < 60''$, except at $\theta \approx 50''$).
  X-ray flux dominated by knots ($\sim 2/3$ of total flux in knots: H03): radio
  surface brightness high: extended $\alpha_{\rm X}$ flat, often
  flatter than $\alpha_{\rm RX}$, and comparable to $\alpha_{\rm X}$
  of knots.

\noindent --- (ii) ($\theta \approx 50''$ and $60'' < \theta <
  190''$). Few X-ray knots ($<10$\% of flux in knots): knots have
  $\alpha_{\rm X}$ less than the extended value. $\alpha_{\rm X} >
  \alpha_{\rm RX}$ and roughly constant, $\approx 1.2 \pm 0.2$. Radio surface
  brightness low, increasing by a factor $\sim 2$ along the jet. X-ray
  surface brightness flat but decreasing by a factor $\sim 2$ along
  the jet. Edge-brightening along the jet is present in the X-ray.

\noindent --- (iii) ($\theta > 190''$; at and beyond the `flare
  point' of H06). $\alpha_{\rm X}$ steepens to very high
  values, $\sim 2.5$. Radio surface brightness increases by factor
  $\sim 3$. X-ray surface brightness
  increases by factor $\sim 3$, then falls off to zero.

In region (i), the flat X-ray spectral index of the extended emission
is consistent with the expectations from shock acceleration coupled
with rapid downstream expansion losses (though some of
the inner knots have well-determined steep spectra, possibly implying
higher magnetic fields in these regions), while the fact that
$\alpha_{\rm RX} > \alpha_{\rm X}$ is inconsistent with a one-zone
model, but is consistent with the idea that much of the particle
acceleration here is localized to small sub-regions of the jet. The
new data provide some evidence that structures that appeared diffuse
in earlier observations (e.g., around knot B) are in fact composed of
smaller sub-clumps. In region (ii), on the other hand, the X-ray and
radio/X-ray spectral indices and the general appearance of the jet --
dominated not by knots but by filamentary, sometimes edge-brightened
features -- are consistent with a truly diffuse acceleration
mechanism, as argued by K06. Finally, as discussed by H06,
region (iii) corresponds to the end of X-ray emission from the jet.
The remarkable X-ray spectral steepening in this region cannot be due
to synchrotron losses -- the spatial scales involved are too large --
but must be due to a progressive loss of efficiency of high-energy
particle acceleration. H06 showed that the mid-IR and radio
data are consistent with the idea that there is no particle
acceleration beyond the end of the X-ray emission at $\sim 240''$: the
new data provide additional evidence that whatever happens at the
`flare point' -- perhaps a significant jet deceleration -- makes the
diffuse acceleration process substantially less efficient at
producing high-energy electrons. Our ability to distinguish between
the different models for the distributed particle acceleration process
(\S\ref{intro}) is limited by the inability of these models to produce
detailed predictions for the particle energy spectrum, but the steep
spectra seen at the end of the Cen A jet represent a challenge for any
current model.

\section{The counterjet}
\label{cjet-section}

\begin{figure}[!t]
\epsfxsize 9cm
\epsfbox{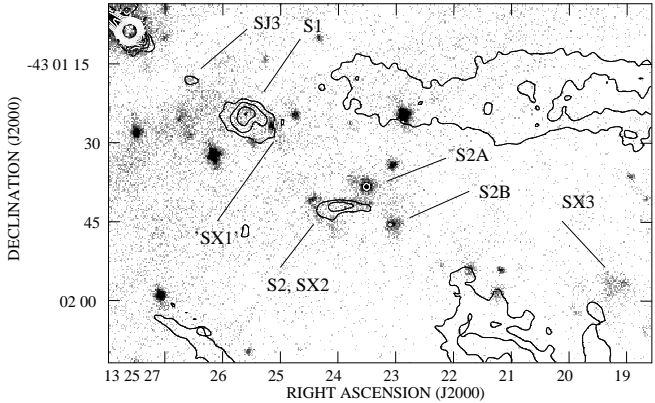}
\vskip -5pt
\caption{The counterjet region of Cen A. Greyscale shows accumulated
  raw counts in the {\it Chandra} observations in the 0.4--2.5 keV
  energy band, binned in $0\farcs 246$ pixels. Contours [$0.1 \times
  (1,2,4\dots)$ mJy beam$^{-1}$] are from the 8.4-GHz
  A+B-configuration VLA map of H03, smoothed to a resolution of
  $1\farcs5$. The radio data are only sensitive to features on scales $\la 20''$.}
\label{cjet}
\end{figure}

H03 showed that a number of X-ray features in the counterjet region
had radio counterparts, proving that a coherent counterjet does exist
on the kpc scale in Cen A, although no large-scale radio emission is
detected from it. With the new X-ray data we detect at least 4 more
counterjet features (Fig.\ \ref{cjet}). There is now a weak X-ray
counterpart to the compact radio knot SJ3 (using the notation of H03)
and to the diffuse feature S1. The feature denoted SX1 by H03, with no
radio counterpart, turns out to be strongly variable (see
\S\ref{variability}) and so may be unrelated to the counterjet.
However, the two radio knots S2A and S2B have very clear X-ray
counterparts with stable X-ray fluxes. There is also a clear
overdensity of fainter compact X-ray features, some of which may be
counterjet knots, though none has a radio counterpart. The most
striking new counterjet features are the {\it extended} features SX2
and SX3, the former of which has a curious V-shaped morphology with a
radio counterpart to one arm of the V, and the latter of which appears
as a patch of diffuse emission on scales of $13''$ (230 pc). Both of
these have clearly harder spectra than the surrounding thermal
emission or than other nearby filamentary structures in the lobe
region (Fig.\ \ref{fc}). We extracted spectra for these regions from
the 2007 datasets only, obtaining a good fit with power-law models
with free absorption as for the other spectra, and found $\alpha_{\rm
X} = 1.42_{-0.14}^{+0.21}$ and $0.85_{-0.15}^{+0.10}$ for SX2 and SX3
respectively, consistent with the range of spectra seen for extended
regions of the jet discussed in \S\ref{spectra}, although we cannot
rule out thermal models with $kT \ga 3.5$ keV. SX3's 1-keV flux
density of $2.3 \pm 0.4$ nJy would imply a 5-GHz flux density for a
counterpart of $\sim 20$ mJy, if it were similar to extended regions
of the northern jet at comparable nuclear distances, which is hard to
rule out given the strong filamentary structure in the S lobe. (For
comparison, 5-GHz flux densities for S1 and S2 are $\sim 60$ and $\sim
80$ mJy respectively.) SX3's similarity in the X-ray to the diffuse
structure of jet features at a similar off-nuclear distance gives us
the first evidence that the diffuse acceleration process can also
operate in the counterjet. If this is so, it is unclear why only this
one purely diffuse counterjet feature is visible in X-rays. The fact
that the jet-counterjet ratio in the X-ray is in general higher for
extended regions than for compact sources lends some support to the
model in which X-ray emitting knots are stationary while the diffuse
emission comes from the bulk flow.

\begin{figure}[t]
\epsfxsize 8.7cm
\epsfbox{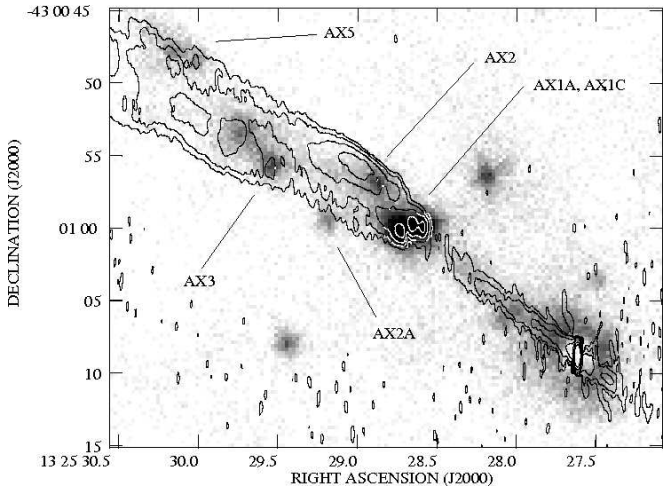}
\vskip -5pt
\caption{The inner jet of Cen A. Greyscale as in Fig.\ \ref{cjet}.
  Contours [$0.15
  \times (1,4,16\dots)$ mJy beam$^{-1}$] are from the 8.4-GHz A-configuration VLA data taken in
  2007, with a resolution of $0\farcs84 \times 0\farcs22$.}
\label{ijet}
\end{figure}

\section{Knot variability and flaring}
\label{variability}

A goal of the Cen A-VLP was to search for extreme variability in jet
knots such as that seen in knot HST-1 of M87 \citep[e.g.,][]{hcbs06}.
Only two bright X-ray features related to the jets show very strong
variability in the ten epochs available. The counterjet feature SX1
appears to have flared in the 2003 observations (H03) when its 1-keV
flux density is $9\pm 1$ nJy compared to $\sim 2$ nJy in all other
observations. Its rather flat spectrum ($\alpha_{\rm X} = 0.4 \pm
0.1$), the fact that it is unresolved and the lack of a radio
counterpart means that it may well be a LMXB unrelated to the
counterjet. More interesting is the feature AX2A detected in the 2007
observations in the inner jet (Fig.\ \ref{ijet}). The 1-keV flux
density of this feature is a constant $5 \pm 1$ nJy in 2007 and $<1$
nJy at all earlier times: it has $\alpha_{\rm X} = 0.65 \pm 0.1$. As
Fig.\ \ref{ijet} shows, it is just at the edge of the radio jet, is
compact or unresolved, and has no detected radio counterpart: given
its flux density, we might have expected one at the level of a few mJy
if it were a jet knot. While continued monitoring of this feature will
be interesting it seems most likely that it too is an unrelated LMXB.
There is thus no evidence for extreme variability in any bona fide jet
knot in Cen A, consistent with the fact that many of the X-ray and
radio knots are spatially resolved. Evidence for weaker variability in
the X-ray and radio will be discussed in a future paper (Goodger \etal
, in preparation).

\section{Summary}

The inner part of the Cen A jet is dominated by the shock-related
knots discussed by H03, although we have so far failed to see the
dramatic variability seen in M87, interpreted as large changes in
particle acceleration processes or in jet flow. The radio-associated
compact features in the counterjet are likely also the result of
small-scale shocks, presumably caused as the counterjet flow
encounters compact obstacles. However, the outer 3.4 kpc of the jet --
and possibly also the newly detected extended X-ray counterjet
components -- are different from the shock-dominated regions in their
X-ray spectra and structure and in their radio/X-ray ratio. Our
results add to the evidence that an unknown, distributed particle
acceleration process operates in the jet of Cen A. Diffuse X-ray
emission with a similar broad-band spectrum is seen in the jets of
many other FRI sources (e.g., 3C\,66B: \citealt{hbw01})
and it also resembles the extended synchrotron X-ray emission seen in
the jets (e.g., K06) and hotspots \citep[e.g.,][]{hck07} of the more
powerful FRIIs. The Cen A data allow us to study the spatial and
spectral properties of this emission and the physical processes
responsible for it in more detail than is possible in any other
object.
\vspace{-10pt}
\acknowledgements
We gratefully acknowledge financial support for this work from
the Royal Society (research fellowship for MJH), the STFC (reseach
studentship for JLG) and NASA (grant GO7-8105X to RPK). The National
Radio Astronomy Observatory is a facility of the National Science
Foundation operated under cooperative agreement by Associated
Universities, Inc.

{\it Facilities:} \facility{CXO (ACIS)}, \facility{VLA}

\end{document}